\documentclass[12pt,a4paper,oneside]{article}
\usepackage{graphicx}
\usepackage{amsmath}
\usepackage{overcite}
\usepackage[sort&compress,numbers,super]{natbib}

\title{\bf 
Molecule-specific Uncertainty Quantification in Quantum Chemical Studies
}
\date{November 9, 2021}
\author{\vspace{0.3cm}Markus Reiher\footnote{contact: markus.reiher@phys.chem.ethz.ch; ORCID: 0000-0002-9508-1565}\\
\textit{ETH Z\"urich, Laboratorium f\"ur Physikalische Chemie,}\\ 
\textit{Vladimir-Prelog-Weg 2, 8093 Z\"urich, Switzerland}\\[2ex]
}

\begin{document}

\maketitle

\renewcommand*{\thefootnote}{\fnsymbol{footnote}}

\begin{abstract}
Solving the electronic Schr\"odinger equation for changing nuclear coordinates
provides access to the Born-Oppenheimer potential energy surface. This surface
is the key starting point for almost all theoretical studies of chemical processes in electronic ground and excited
states (including molecular structure
prediction, reaction mechanism elucidation, molecular property calculations, quantum and molecular dynamics).
Electronic structure models aim at a sufficiently accurate approximation
of this surface. They have therefore become 
a cornerstone of theoretical and computational chemistry, molecular physics, and materials science.
In this work, we elaborate on general features of approximate electronic structure models such as
accuracy, efficiency, and general applicability in order to arrive at a perspective for future developments,
of which a vanguard has already arrived. 
Our quintessential proposition is that meaningful quantum mechanical predictions for chemical phenomena
require system-specific uncertainty information for each and every electronic structure calculation,
if objective conclusions shall be drawn with confidence.
\end{abstract}

\newpage

\section{Introduction}
Electronic structure theory\cite{Helgaker2000,Parr1989}
is at the core of many theoretical and computational approaches in
the molecular sciences as well as in condensed matter and solid state physics. 
The Born-Oppenheimer approximation \cite{Tully2000} 
introduces the electronic Schr\"odinger equation that treats all electrons of an atomistic system as dynamical
entities moving in the field of a fixed nuclear scaffold. The success of computational molecular and materials science
is rooted in this framework, and naturally, an ever growing number of electronic structure models has been devised in the 
past decades that aim at improving on existing approximations in terms of general applicability,
increased accuracy, and optimized efficiency with respect to computational time.
The rate of introduction, adoption, and use of capital-letter acronyms that denote the various models and approximations 
may serve as a good measure of these efforts.
In this work, we assume a bird's eye view on the current status of developments 
in an attempt to assess and determine feasible options for future models.
Before we provide our perspective on these developments, we first dwell on the theoretical foundations
of electronic structure models.

\section{Theoretical Foundations}

The explicit decoupling of nuclear and electronic
degrees of freedom through the Born-Oppenheimer approximation
is, in many cases, an efficient and reliable approximation, but most importantly, it can be systematically
improved to include nonadiabatic and rovibronic couplings.\cite{Tully2000} 

The Born-Oppenheimer approximation introduces the concept of molecular structure as a spatial arrangement of nuclear positions,
making a local minimum structure the representative of a stable compound.
The emergence of (classical) molecular structure can be studied within a full quantum framework (see Refs. 
\citenum{Woolley1976,Matyus2011,Matyus2011a,Muolo2019a}
and references therein), 
but the application of such a general pre-Born-Oppenheimer approach for 
calculations on large systems is unfeasible in practice. The reason for this is not only the prohibitive computational cost of
a truly accurate all-particle quantum calculation, but also
the lack of understanding when the concept 'potential energy surface', i.e., the change of the electronic
energy with nuclear coordinates, is simply not available. Instead, in pre-Born-Oppenheimer quantum theory,
we obtain only a spectrum of (molecular)
energy eigenvalues for the total system of electrons and nuclei without
any direct relation to molecular structure. For this reason, electronic structure theory in the Born-Oppenheimer
approximation has become the prevailing theoretical approach because it allows us to think in terms of classical molecular structure
and to improve on this approximation a posteriori in an efficient manner.

From the perspective of practical considerations and
numerical calculations, the Born-Oppenheimer approximation has therefore become the key entry point for actual calculations of
molecules. 
Its most decisive consequence is the assignment of an energy, the electronic energy, to a given molecular
structure -- which may be an equilibrium structure of a chemical compound or an arbitrarily distorted nuclear scaffold
-- by virtue of the emerging electronic Schr\"odinger equation (contributions of the nuclear degrees of freedom can
be included subsequently). 

The decisive difference of quantum chemical models compared to other types of atomistic modeling approaches -- such as force fields that govern the 
interactions of atoms (as new emergent entities, rather than electrons and nuclei) -- is that the interaction of their dynamical objects, i.e., the electrons and nuclei, 
is well known and determined by the laws of quantum electrodynamics. If the speed of light is taken to be infinite, then the
interaction law will be known exactly and simply be given by the electrostatic Coulomb interaction of two point charges promoted to an operator,
whereas magnetic interactions and retardation effects vanish in this limit \cite{Reiher2015}. This advantage together with the fact that
any atomistic system is built of electrons and atomic nuclei makes quantum chemical models universally
applicable. Hence, these models that rest on the first principles of quantum mechanics enjoy a special status in theoretical
chemistry and molecular physics. 

It is often taken for granted that successful electronic structure models can be applied to arbitrarily sized molecules (of course,
up to a point
where the size of the molecule will be in conflict with computational feasibility). In other words, if the size of a molecule under study
changes (e.g., by a substitution of a residue, for which benzene and toluene may be taken as an example), then we will still expect our computational
models to work if we start the computer program that calculates an approximation to the electronic energy. 
Hence, an implicit requirement for the development of general approaches 
designed to solve the electronic Schr\"odinger equation has been that they can
be applied efficiently for a fixed but otherwise variable number of electrons; only the scaling behavior of an approximation
with system size will put some upper limit on this generality depending on the actual computer hardware available.

Irrespective of available hardware, very accurate electronic structure 
methods scale unfavorably with system size and therefore require a significant computational effort.
As a consequence, they can only be applied to far smaller systems than those accessible to more approximate approaches,
which trade accuracy for applicability.
Although system size is a moving target in view of continuously evolving
algorithmic improvements and hardware capabilities, the situation does not change in principle: 
very large systems are approached with rather approximate
methods, whereas very accurate methods remain to be bound to very small molecules.

It is mostly for historical reasons that all electronic structure models have been first devised for the ('non-relativistic')
Schr\"odinger equation rather than for many-electron Hamiltonians based on Dirac's theory of the electron, which delivers
an equation of motion that is Lorentz covariant for a single electron in a classical electromagnetic field.\cite{Reiher2015} Considering the fact
that observables such as energies are the prime target quantity of computations, 
the fundamental Einsteinean framework of special relativity 
requiring Lorentz invariance of all fundamental physical equations has been reduced
to its numerical effect on observables obtained by relativistic quantum chemical methods. 
For example, Taylor series expansions in powers of the inverse speed of light
have been considered early on and introduced new concepts such as spin-orbit, spin-spin, spin-other-orbit interactions, and
kinematic relativistic effects. Although there exist convergence issues with this type of Taylor series expansions,
they deliver well-defined analytical expressions for operators that can be used in perturbation theory to
correct the electronic energy for such effects. Naturally, efficient variational relativistic procedures have also been devised.\cite{Reiher2015}

Important for the assessment of errors of approximate relativistic energy
operator expressions is that the dependence of relativistic approximations on nuclear charge, electron kinematics, and spin   
appears to be such that no appreciable
error accumulation or growing uncertainties are found in 
relative quantities such as reaction energies. This is due to the fact that relativistic effects are often
atomically conserved (e.g., during reaction) and therefore remain largely unchanged in chemical transformations.
This is also the reason why surrogate potentials that model relativistic atomic cores (e.g., effective core
potentials and pseudo potentials \cite{Dolg2012}) are reliable in capturing relativistic effects (and even reduce the computational effort 
through elimination of core electrons and alleviate the discretization error of a finite basis set through elimination
of the nodal structure of the wave function in the core region). 

For what follows below, the most important consequence is that no (approximate) relativistic method will introduce an
error in the electronic energy that will be difficult to control when the underlying nuclear scaffold changes (e.g., during chemical
reaction). In other words, the change of electronic energy due to the principles of Einstein's theory of special relativity
and quantum electrodynamics
is very well understood, up to the point where it has become possible to calculate the binding energy of molecular hydrogen
to an absolutely remarkable precision \cite{Puchalski2017,Puchalski2018,Puchalski2019,Wang2018}, in excellent agreement with high-precision spectroscopic measurements.\cite{Cheng2018,Holsch2019} 
Effects that transgress the realm of quantum electrodynamics (such as electroweak effects \cite{Berger2000}) are known to
be orders of magnitude smaller. 

\begin{center}
{\bf Statement 1:} 
{\it
The theoretical foundations of electronic structure theory are very well established and understood.
}
\end{center}

Various well-defined relativistic generalizations of the 
standard non-rela\-ti\-vistic electronic Schr\"odinger equation have been developed, which connect the quantum theory
for atomistic systems to the fundamental theory of quantum electrodynamics.\cite{Reiher2015} Since these relativistic variants do not pose
any unsurmountable hurdle to approximations designed for the efficient solution of the non-relativistic electronic Schr\"odinger equation, 
electronic structure methods developed in the non-relativistic regime
apply equally well in the relativistic domain.
By contrast, electronic structure theory has been plagued by
the peculiarities of many-electron systems in general, which are first and foremost the Coulomb interactions of the (point) particles
featuring a singularity where the particle coordinates coalesce.

\begin{center}
{\bf Statement 2:} 
{\it
The quantum mechanical Coulomb interaction of two electrons is the crucial challenge for accurate approximations.
}
\end{center}

\section{Electronic Structure Models}

Two main classes of approximations have emerged in electronic structure theory: wave-function-based theories 
\cite{Helgaker2000} and density functional theory (DFT) \cite{Parr1989}, with variants that
bridge both classes. In actual calculations, DFT is often applied as a parametrized model because of 
approximations in the exchange-correlation energy functional that require parameters, in most cases determined for some
reference data. 
Despite the universal nature of these parameters,
which are taken to be valid for any sort of atomistic system,
and despite the first-principles nature of DFT per se, actual applications are sometimes classified as 'semi-empirical' refering
to the parameter dependence on the reference (training) data and to the, to a certain degree arbitrary, 
assumptions on the analytical form of
a functional. 
Proposals to improve on this situation have been numerous,
but the general dilemma of electronic structure theory, i.e., the ineluctability to 
compromise accuracy for computational efficiency, also hampers these attempts. This is already seen in the standard
approximations of DFT, which Perdew ranked as a Jacob's ladder \cite{Perdew2001} (a hierarchical ranking that has been confirmed
in rigorous and extensive numerical studies \cite{Mardirossian2014a,Mardirossian2017,Mardirossian2018}), for which improved accuracy comes at significantly increased computational cost.

Although the existence of the exact exchange-correlation energy functional has been proven,\cite{Hohenberg1964,Lieb1983} 
its exact analytical form is and most likely will remain unknown. In view of the increasing complexity of advanced density
functionals,\cite{Becke2014} one is struck by the impression that the exact exchange-correlation functional (or a very accurate approximation
to it) will be as hard to evaluate as a wave-function model of the same accuracy.

By contrast, wave function theories often allow for systematic improvement owing to a systematic and well-controllable
construction of the electronic wave function through basis-set expansion techniques -- at least in principle, but
not to an arbitrary degree in practice. In a mathematically stringent way, the electronic wave function can be
expanded in terms of a complete many-electron basis set. The
many-electron basis functions can be systematically built (usually as determinants) from products of one-electron functions
(e.g., Hartree-Fock orbitals or some general type of energy-optimizing orbitals),
which themselves can be generated as a linear combination of pre-defined and therefore known one-electron basis functions 
(representing the well-known linear combination of atomic orbitals).

All standard models -- such as
configuration interaction, coupled cluster (CC), complete active space self-consistent field, or M{\o}ller-Plesset perturbation theory --
approximate the electronic wave function by a superposition of these many-electron basis functions, of which an ordered 
set is
generated and classified according to orbital substitutions compared to a (dominant) reference determinant 
(so-called 'excitations'). Although this choice of singles (S), doubles (D),
triples (T), ... excitations for the definition of classes of many-electron basis functions 
produces a well-defined approximation hierarchy, it does, a priori, 
not guarantee optimum convergence as it is not clear which
'excitations' will be decisive for the model to achieve a given accuracy. However, as long as changes in molecular structure
are local (as is the breaking of a chemical bond in a chemical reaction), very high excitations may be avoided.
A vast body of numerical results has therefore led to the wide acceptance of the CCSD(T) coupled cluster model \cite{Raghavachari1989}
as a 'gold standard' for quantum chemical
calculations. It is, however, important to emphasize that, although this approach can deliver results 
of chemical accuracy of about 1 kcal/mol in benchmark calculations,
the precise accuracy of the approach for some given case is, in general, not known. Instead, one resorts to ensure some general
constraints on the electronic structure of a molecule under investigation that must be fulfilled in order
to achieve a high accuracy with the CCSD(T) model. These constraints require a (close to) equilibrium molecular structure, a closed-shell
electronic structure, absence of multi-reference effects that are typical for open shells and so forth.

Unfortunately, no such model of similar reliability, but universal applicability could be established for the 
general multi-configurational case. Although 
many open-shell and bond-breaking situations may be well captured by ground-state coupled cluster theory with
a not too high degree of excitations in the electronic wave function, 
it is troublesome that the most general case has escaped attempts to produce an
equally accurate and broadly applicable electronic structure model.
Surely, modern complete-active-space approaches such as the density matrix renormalization group (DMRG) \cite{White1992,White1993,Baiardi2020} or full configuration
interaction quantum Monte Carlo (FCIQMC) \cite{Booth2009,Ghanem2019,Ghanem2020} have pushed the limits to a degree where any (for reaction chemistry) relevant active orbital
space can be treated (i.e., of up to one hundred active spatial orbitals). However,
the neglected dynamic correlation from the huge number of weakly correlated 
orbitals, which are outside the active space, escapes a treatment on the same level of accuracy; typically
only multi-reference perturbation theory to second order represents the highest level of accuracy achievable.\cite{lind20} 
Continuous efforts have tried to improve on this situation; examples are
tailored coupled cluster,\cite{Bartlett2005,Veis2016,Morchen2020}
combinations with short-range DFT \cite{Savin1996,Savin1997,Jensen2007,Hedegard2015,Jensen2018,Rodriguez-Jimenez2021,Pernal2021}
or transcorrelation \cite{Boys1969,Shiozaki2012,Luo2018,Baiardi2020a,Guther2021}
to treat the electron-electron cusp due to the singularity of the Coulomb interaction, 
to mention only a few. 
Already the amount of suggestions for accurate multi-configurational methods (not reflected in the list of references of this work
and beyond its scope) may be taken as an indication
that no universally satisfactory and generally accepted best solution has been found so far.
At the same time, multi-reference coupled cluster theory has suffered from formal difficulties,\cite{Evangelista2018} by which each of the various
ans\"atze is plagued to a different degree. 
It is therefore no surprise that other advanced configuration interaction (CI) methods have been developed in recent years 
\cite{Umrigar2016,Whaley2016,Evangelista2016,Zimmerman2017,Zimmerman2017a,Eriksen2018,Loos2018,Chilkuri2021,Chilkuri2021a}
as a potential way to approach full configuration interaction (FCI) results.

Hence, one may doubt whether this situation can in principle be improved on, and a quantum leap to solve these principal problems,
that will also need to stand scrutiny in terms of computational feasibility, appears not very likely on traditional computers. 
Naturally, many have wondered whether there are other, better suited approaches that await discovery or
just need to flourish.\cite{Popelier2011} Hence, alternative approaches have been sought for, but all developments rely on
more or less the same principles and technologies and still
wait for an eventual breakthrough. Even if this happens in the future, history teaches us to expect that there will 
always be strings attached that will eventually restrict the range of applicability and the achievable accuracy.
And therefore, the basic train of thought of this work will persist.

The most dramatic improvement may come from hardware development in the future, 
of which the most prominent example is the hope to solve electronic structure problems with quantum hardware to be developed,\cite{Abrams1997,Abrams1999,AspuruGuzik2005,McArdle2020,Liu2021}
which may deliver FCI-quality results up to a pre-defined accuracy by quantum phase estimation (QPE). Although current implementations
rely mostly on the variational quantum eigensolver,\cite{Peruzzo2014} QPE can be shown to be feasible on moderately sized future quantum
computers for actual chemical (rather than toy) problems.\cite{Reiher2017,vonBurg2021}
In this context, it is important to understand that a typical electronic structure problem represented
in an orbital basis of, say, 1000 one-electron basis functions, will require 2000 logical qubits to represent the
state (which, in turn, will require orders of magnitude more physical qubits for error correction purposes), let alone the 
problems of initial state preparation and loading the Hamiltonian parametrization in second quantized form into
the machine (consider, in particular,
the two-electron integrals for an active space of 1000 orbitals). However, if such large universal digital quantum 
computers become
available in the future and if they are widely accessible, 
then they can produce results of FCI quality for molecular sizes that can accurately describe
any (usually local) chemical reaction event, which will render very many traditional electronic structure models superfluous
(for instance, the separation into static and dynamic correlation may become unnecessary).

As a side remark, the ongoing efforts in establishing a universal quantum computer show that
one must be willing to cool down a quantum computer 
to very low temperature. However, such a strategy will also be beneficial for traditional approaches directed
towards cryogenic computing.

For the reasoning in this paper, we adopt the point of view that 
we will have to cope with the situation as it presents itself to us today
in the near and maybe also in the distant future. Namely,
that we will accomplish incremental improvements in all directions, but that no generalist approach of arbitrary accuracy
for some system under study is likely to emerge.

\begin{center}
{\bf Hypothesis:}
{\it
 The current situation of electronic-structure models might represent the end point of our efforts to
devise universally applicable, accurate, and feasible electronic structure models. In view of
the tricks and techniques used in all developments, it appears to be unlikely that paradigm-shifting new ideas will emerge
in the (near) future in view of the fact that all viable approaches eventually rely on some version of the configuration interaction
expansion.
}
\end{center}

Even if this hypothesis will be falsified in the future, it is the best starting point for a discussion on how
to deal best with our current arsenal of electronic-structure methods for solving pressing chemical problems.

\section{The Role of One-Electron Basis Sets}

It is hard to obtain accurate quantities for many-electron problems. Already
the choice of a typical one-electron basis set usually affects the total electronic energy
significantly, because basis sets are designed for specific purposes and compromise on others for the sake of
computational feasibility. For instance, to allow for sufficiently accurate
results of relative electronic energies for chemical reactions, 'atomic orbital' basis sets attempt to represent the
valence-shell region well
by suitable polarization and diffuse functions. The core shells, i.e., those molecular orbitals that appear like
atomic $1s$-type orbitals, are not very well represented, which affects their orbital energy. If one expresses the
total electronic energy as a function of orbital energies (e.g., as the sum of all canonical orbital energies plus the one-electron
integrals in Hartree-Fock theory divided by two) then it is obvious that the large core-shell orbital energies affected by
a large error propagate this error to the total electronic energy. However, for relative electronic energies of processes where
atomically conserved features of the electronic structure are unimportant, these errors will drop out as they are conserved 
in reactants and products.

Very many numerical studies have been conducted to demonstrate the role of the various ingredients (including 
quasi-relativistic Hamiltonians) of which we may refer to three studies presented in Refs. \citenum{fell06,fell07,pete12}.
Apart from the electronic energy, also calculations of its derivatives -- which define molecular properties through
response theory \cite{Norman2018} and can be taken as 'relative quantities' by definition (as is particularly obvious in the finite-field
approach) -- usually factor in error compensation, but may require special functions to represent
a property well. For instance, polarizabilities naturally require one to allow for polarization effects, and hence,
polarization functions will be very important.\cite{Sadlej1988} Taking further derivatives, e.g., for the calculation of
Raman and Raman Optical Activity spectra has led to the proposal of tiny skeleton basis sets with polarization functions
to reduce the computational effort \cite{Zuber2004} which turn out to be quite accurate for these 'relative' quantities.\cite{Reiher2005a}

The singularity at the coalescence of the electron-electron Coulomb interaction will demand extended
one-electron basis sets, especially for correlated electronic structure methods. To achieve chemical accuracy with decent triple-zeta
basis sets requires the introduction of explicitly correlated functions, as elaborated into the now established F12 ansatz,\cite{Ma2018} 
or a modification of the Hamiltonian as in the transcorrelation approach \cite{Boys1969} or through range separation
by short-range DFT.\cite{Savin1996}

In general, even large orbital basis sets therefore suffer from errors that are well above 1 milliHartree, an accuracy that nevertheless
may be approached for {\it relative} quantities due to the fortunate error compensation of atomically conserved parts of electronic
structures. While this observation has been central to the application of electronic structure models in computational quantum chemistry,
rigorous mathematical results on the discretization error introduced by finite basis sets are still needed (see Ref. \citenum{Cances2017a}).

Large standard valence-optimized one-electron basis sets 
will not allow one to reach a high accuracy in the total electronic energy of, say, 1 nano\-Har\-tree,
for which one is advised to switch to explicitly correlated basis functions such as geminals.\cite{Adamowicz2013,Matyus2012}
Depending on the target problem, such an accuracy might be needed. While kinetic details of a reaction will require one to
know the relative electronic energy to an accuracy better than 1 kJ/mol, which is somewhere inbetween milliHartree and microHartree 
accuracy, nanoHartree accuracy may become important for high-resolution spectroscopy, of which the measurement of the
H$_2$ binding energy is a prime example.\cite{Puchalski2019,Wang2018,Cheng2018,Holsch2019} 
It is therefore obvious that {\it the required target accuracy} needs
to be factored into the effort one spends on solving the electronic structure problem.

\begin{center}
{\bf Statement 3:}
{\it
 While the discretization error introduced by orbital and geminal basis sets -- as well as any other error that results
from the technical implementation of solution procedures -- may be systematically reduced, its
value in an actual calculation is usually not precisely known.
}
\end{center}

\section{Can We Know the Error by Which a Specific Quantum Chemical Result is Affected?}

The Ritz variational principle provides us with a means to judge the reliability of total electronic energies
obtained with approximate (variational) electronic structure models by assuring that they are
upper bounds to the ground-state energy eigenvalue of the electronic Hamiltonian.
As a consequence, we know that the lowest approximate energy represents the lowest upper bound and, hence,
the best approximation to the exact total electronic
energy. To assess the remaining error requires, however, a lower bound that can be efficiently computed. Despite continuous development
devoted to establishing such rigorous lower bounds for Hermitian operators
(see Refs. \citenum{Cances2017,Cances2018,Martinazzo2020a} and references therein), this ansatz remains underexplored. 
Moreover, even if such a lower bound can be efficiently computed, this does not imply that the 
uncertainty obtained as the difference between both bounds will be sufficiently small to allow for 
a tight bracketing of the error so that it is eventually useful.

Analytical error assignment based on the structure of the underlying differential equations (i.e., the electronic
Schr\"odinger equation or its descendants, the self-consistent field equations) is very hard as well. 
The external potential created by the nuclear scaffold presents a complicated variety as a coefficient function in
the partial differential equations. At the level of total electronic states, compared to some reference molecular structure, slightly shifted nuclear positions 
create a shifted energy eigenvalue spectrum. This can then represent a very different electron correlation pattern and 
may even give rise to emerging avoided crossings or
conical intersections.\cite{Domcke2011} At the orbital level, the fact that construction of the electron-electron interaction 
requires knowledge of the orbitals themselves -- which is dealt with in the self-consistent field procedure --
makes the corresponding one-electron partial differential
equations hard to analyze.

A change in the external potential (e.g., as it occurs along a reaction coordinate) is continuous and
typically changes the error of an electronic structure model in a smooth and continuous, non-abrupt fashion (even if this error may,
in some cases,
successively become very large as, for instance, in coupled cluster calculations of at most doubles excitations in
double-bond breaking processes). As such, the error is systematic and, obviously, it does not depend on any sort of noise
(as long as the computer hardware is reliable, which cannot be strictly guaranteed.\cite{hoch21}) However,
this usually does not allow one to explicitly know the error in a specific quantum chemical calculation
without carrying out an improved calculation. Unfortunately, the latter is, for various reasons, typically unfeasible
(otherwise one would have chosen the more accurate (and, hence, more expensive) calculation from the start).

\begin{center}
{\bf Statement 4:} 
{\it
The individual absolute error of a specific quantum chemical result is usually very hard,
if not impossible, to assess accurately. 
}
\end{center}

\section{Systematic Improvability}

For the purpose of this work, it is instructive to now look at electronic structure methods from the point
of view of how errors introduced by various approximations are actually assessed in routine calculations.

It is the specific nature of first-principles methods to be well-defined in terms of the fundamental
interactions requiring hardly any model assumptions on the interaction operators -- other than those that emerge from quantum electrodynamics.
This formally convenient and intellectually appealing situation has given rise to the romantic belief that a 
general electronic structure model -- applicable to any atomistic system with sufficient accuracy 
(say, with chemical accuracy of at least 1 kcal/mol) and feasible -- can actually be devised if we only search for it with sufficient
devotion and dedication. Because of this attitude that we may be able to approximate the exact solution of the electronic
Schr\"odinger equation with sufficient accuracy, the modeling of electronic structure has hardly been scrutinized by 
rigorous uncertainty quantification procedures applied routinely to specific atomistic systems under investigation.

It is natural for many-particle basis-set expansion approaches such as configuration interaction and coupled cluster theories that one
can approximate the exact solution of the electronic Schr\"odinger equation to arbitrarily high accuracy under the provision 
(i) that one ignores the exponentially scaling wall created by the exploding number of many-electron basis functions
with system size (even when acknowledging that there may be a lot of deadwood in many-electron Hilbert space that 
might be avoided by smart procedures, such as DMRG and FCIQMC mentioned above) 
and (ii) that one has chosen some suitable finite one-electron (orbital) basis set in which an accurate solution can
be obtained. Following the excitation hierarchy in these schemes allows one to systematically 
approach a 'complete' (finite) many-particle basis set and, hence, the full configuration interaction
energy in this basis. 

Interestingly, this idea of systematical improvability is hardly pushed to the limit in practical calculations,
where one prefers to pick just one well-established generalist model (such as CCSD(T)) to have only one calculation
to rely on. Only then, one does not drown in numerical data, can focus on the scientific question to be answered, and does
not spend additional computational time on it. However, in such cases it is not at all clear how reliable the calculated data
are for answering a chemical question (regarding, for instance, the elucidation of mechanistic pathways).
While the prevailing assumption is that CCSD(T) combined with a suitable one-particle basis set will be sufficiently
accurate, this situation actually points toward the need for system-specific
error assignment with little additional computational effort.

Pople and co-workers arranged wave-function-theory models according to their expected accuracy, in order to establish
extrapolation schemes that allow one to exploit orbital bases of different size to make the high-excitation-degree wave functions
accessible, leading to combination theories such as G2 \cite{Curtiss1991,Curtiss1992,Curtiss1995} and G3 \cite{Curtiss1998,Curtiss1999,Curtiss1999a}. Such composite models have been further developed by other groups
(see, for instance, Refs. \citenum{Martin1999,Boese2004,DeYonker2006,Karton2006,Barnes2009,Das2010,Alsunaidi2016,Karton2017} to mention only a few).
At the same time, the result of such approaches is still
affected by an error, even if it may be significantly reduced, which is not known precisely for a specific system under study
(despite recent efforts to equip composite protocols with uncertainties for specific quantities and pre-defined classes of molecules.\cite{bako19,bako20})

In DFT, the situation is more cumbersome because of the more or less empirical introduction of parameters and analytical functional forms that define
the approximate exchange-correlation energy functionals. They
do not allow for a systematic improvement as one has no idea about what to change in order to
arrive at a more accurate answer. In other words, having obtained a DFT result
with some density functional, by and large, presents us with the problem that it is not clear 
how to improve on it or even how to scrutinize it. Surely, climbing up
the notorious Jacob's ladder \cite{Perdew2001} could be one option, but this concept is more heuristic than rigorous at its core.
In particular, improved results can only be expected in an average manner,\cite{Mardirossian2017} whereas a specific result for a 
specific system may not improve even if one takes the next rung of the ladder.

\begin{center}
{\bf Observation:}
{\it
 Electronic structure models are affected by some error that is usually not assessed in a specific application.
Instead one often relies on experience and intuition gained with some approach.
Most approximations rely on error compensation that leads to more reliable results for 
relative quantities, but their precise accuracy for a specific case under study is typically not known either.
}
\end{center}

Whereas one may quote very many examples for this observation,  
which would be far beyond the scope and purpose of this
work, we may instead take it as a stepping stone in our gedanken experiment that weighs in on the future 
of electronic structure methods. We may therefore continue our discussion from this conclusion:

\begin{center}
{\bf Conclusion 1:}
{\it
 While we have a very good understanding of what approximations in electronic structure theory 
are feasible and efficient, they are, in all cases, 
affected by an error that will be unknown for a specific molecular structure under consideration, even if this
error is expected to be small for certain approaches.
}
\end{center}

\section{Traditional Error Assessment Strategies} 

Rigorous error estimation approaches based on reference results are not without difficulties. 
The generalist approach of electronic structure has relied on representative
benchmarking, where reference or training data have been selected based on 
chemical intuition or simply based on availability. 
As a result, transferability -- usually implicitly assumed if mean absolute deviations are found to be sufficiently
small for training or test data sets -- may not come without accepting the fact
that one may encounter large unexpected errors. Large absolute deviations 
obtained for reference data might differ significantly from the mean
absolute deviation, which may be taken as a first hint of compromised transferability. 
However, these standard measures 
can only provide indications
about errors to be expected for test data if the reference data can be shown to be related in some way to the
test data. The latter condition is not trivial to demonstrate and certainly not a question of experience
or chemical intuition.
Moreover, only few more sophisticated ways have been proposed to improve on the stochastic insight gained for
benchmarking against a specific test data set.\cite{Proppe2017a,Pernot2017,Pernot2017a,Weymuth2018,Pernot2018,Pernot2020}

A major problem has been that, historically, reference data sets have not been rigorously 
developed based on stringent stochastic criteria, but often emerged 
as being determined by availability of reliable data rather than by demand of specific data. 
The general situation improved in the new millenium,
but accurate reference data is still lacking for challenging problems such as transition metal complexes 
or (very) large molecules. Despite meritorious efforts to enlarge these reference sets (see, for example, Refs.
\citenum{Korth2009,Goerigk2010,Goerigk2017,Mardirossian2017,Dohm2018})
in order to cover typical chemical phenomena with data points that appear to be reasonable. 

While the focus here is mostly on the primary quantity obtained in an electronic structure calculation, i.e., the
electronic energy, we emphasize that other molecular properties and atomistic decompositions thereof have also been the subject of
detailed benchmarking studies (for two examples see Refs. \citenum{baus87,matt10}).

A main problem which remains is that parametrized models are usually not scrutinized by expansive statistical means. 
Surely, the adoption of a parametrized model introduces an error that cannot be overcome within this model, 
but fixing the model parameters at some reference data may be affected by biases that have gone unnoticed.
It is for this reason that we have advocated\cite{Proppe2017a,Simm2017b} 
basic standard tools such as bootstrapping \cite{Efron1979} and jackknifing \cite{Miller1974} for
the elucidation of confidence intervals and the assessment of the importance
of individual data points, respectively, in quantum chemical calculations. 

With bootstrapping and jackknifing, we were able to demonstrate,\cite{Weymuth2018} for instance, that the error
development of an increasing number of pairwise semiclassical dispersion corrections may not necessarily lead
to error compensation, but instead to error accumulation. While this is true for the absolute value of the total
dispersion energy, also relative dispersion energies can be affected by error accumulation and blindly trusting in 
error compensation in comparisons may be too risky.
These findings may cast some doubt on the general transferability of such corrections to large molecular systems,
whose size transgresses that of those molecules for which the corrections were parametrized. 

More importantly, results obtained for benchmark sets suffer from the fact that one does not know to what degree they
are transferable to an actual case at hand. 
Surely, how severe this transferability issue is will also depend 
on the required target accuracy in such a way that a tolerable low accuracy may alleviate transferability concerns.
Still, one will require some objective measure by which the actual accuracy can be assessed or, at least, estimated.

Transferability of a given electronic structure model to a specific new atomistic problem is 
not guaranteed, and unexpectedly large errors can be encountered, often without warning (for examples see
Refs. \citenum{Reiher2001,Salomon2002,Cohen2012,Weymuth2014,Liu2015,Husch2018}).
These encounters of unexpected outliers may be taken as an indication that 
reference data are likely to be incomplete if chemical accuracy shall be reached for arbitrary systems 
and point to the need of continuous benchmarking monitored by an uncertainty quantification protocol.
It is even questionable whether we will ever have enough benchmark data at our disposal that allow us to show that
transferability to unseen molecular systems is guaranteed in a way that a certain small error margin is not exceeded,
ignoring the fact that this would also require one to define a reliable measure of transferability.

\begin{center}
{\bf Corollary:}
{\it
Transferability of benchmark results to a specific case under investigation cannot be guaranteed in a rigorous manner. 
}
\end{center}

\section{Bayesian Uncertainty Quantification} 

A crucial conclusion for and from numerical studies 
with various electronic structure models is that the exact error of a given model 
for a specific molecule is not known. For instance, if the largest absolute deviation obtained for a model on some benchmark data
turns out to be much larger than what would be tolerable for drawing sufficiently reliable conclusions from the calculated
data, no guarantee can be given
that calculated results will be sufficiently reliable for the specific problem. For most electronic structure
models, this actually implies that system-specific uncertainty quantification will be imperative, if predictive (de novo) work
shall be accomplished -- even if chemical accuracy, which does not pose a very tight constraint compared to spectroscopic 
accuracy, tolerates certain errors.
If one follows this line of thought, the next question is how uncertainty quantification can be conducted in a way that
is reliable and does not come with a computationally prohibitive overhead.

In a theory proposed by Sethna and co-workers \cite{Mortensen2005} for DFT calculations, the sampling error of a class of
functionals is mapped onto the real error at a benchmark data set in such a way that one can exploit the sampling of calculated
results for error estimates for systems unrelated to the benchmark set. Unfortunately, the mapping affects only the linear
parameters in a functional.\cite{Simm2016} Nevertheless, this Bayesian approach that transfers knowledge obtained for benchmark
data to unseen molecular structures led to the development
of Bayesian error estimation functionals.\cite{Wellendorff2012,Wellendorff2014} Unfortunately, it turns out that mean and largest absolute deviations
remain comparatively large for this class of functionals.\cite{Wellendorff2012,Wellendorff2014}
Concerns have been raised that the parametric uncertainty does not necessarily have the same 
functional shape as the model error's amplitude in general.\cite{Pernot2017,Pernot2017a}

In principle, a system-focused approach 
that re-adjusts the non-linear parameters to system-specific benchmark data achieves a significantly increased accuracy.\cite{Simm2016} 
Still, even for such a system-specific
optimization of a state-of-the-art density functional, the error estimation turns out to be not flawless on the training data, 
i.e., it does not reproduce the known error bars in all cases.\cite{Simm2016}
Hence, even a system-specific approach suffers from the fact that the analytical form of the density functional is not flexible 
enough. Another issue with this methodology is that it 
requires extensive DFT recalculations when benchmark data is extended for a given functional.

Acclaimed machine learning models \cite{rupp15} may step in and provide an alternative approach with various benefits. 
For one, they can provide
a glue between data of different origin. Hence, they can connect data obtained with a fast-but-inaccurate model to those of
an accurate, but computationally more expensive model. In this way, they even access the model-inherent error of a
parametrized model (such as DFT or semiempirical methods) compared to an accurate model with less parameters or assumptions
(such as explicitly correlated coupled cluster theory).
Moreover, specific machine learning models such as Gaussian process regression (GPR) come 
with built-in uncertainty quantification that allow one to rate the results obtained. 
However, any machine learning model may be straightforwardly equipped with suitable statistical measures such as
standard deviations and confidence intervals, even neural networks.

Since machine learning approaches are data driven, they rely on reference data and 
provisions must be taken for their production (see below).
As such, they are Bayesian in their very nature, and therefore, these protocols
are not easy to accept from a puristic first-principles point of view,
in which one attempts to approach the exact solution to a problem -- at least in principle.

For example, we have demonstrated how to exploit machine learning 
to improve on semiclassical dispersion corrections 
(in this case by GPR) 
\cite{Proppe2019a} and on a classical force field.\cite{Brunken2020} In these cases, the existing model remains intact and 
the machine learning approach corrects for deficiencies in this fixed model. The idea behind this procedure is 
that the physical model may capture much of the physics of the problem, leaving only deficiencies for the machine learning model
to learn and correct so that, ideally, less reference data will be required in such a $\Delta$-machine learning \cite{Ramakrishnan2015} type of approach.

Machine learning models are data driven, but agnostic with respect to the type of data. As an example, we note that
GPR has been successfully applied to exploit system-specific information on confidence intervals for the 
optimization of molecular structures.\cite{Raggi2020} 

We also exploited machine learning (again GPR) to assess the accuracy of approximate electronic
structure models deployed in the elucidation of large amounts of related molecular structures (e.g., encountered in 
reaction-network \cite{Sameera2016,Vazquez2018,Dewyer2018,Simm2019,Unsleber2020} and 
chemical-space exploration \cite{Shoichet2004,Pyzer-Knapp2015} or first-principles molecular dynamics \cite{Marx2009}) 
by comparing to more accurate reference
data. Here, the key advantage of GPR is to determine for which structure a reference calculation is required.\cite{Simm2018}
This is encoded in
a measure of molecular similarity, for which many inputs are available such as the Coulomb matrix \cite{Rupp2012}
and the smooth overlap of atomic positions \cite{Bartok2013}. However, more work on such measures will be required as they 
also need to be faithful for relating very similar molecular structures such as an educt and product connected by a transition
state structure. Here, the former two may be affected by similar errors in the approximate method while the error
can be larger for the transition state structure, although it is more similar to educt and also to product than educt and
product to one another. 

Our approach\cite{Simm2018,Proppe2019a,Brunken2020,Brunken2021} has two key advantages: It is computationally cheap to backtrack new benchmark data which were determined
by the machine learning model, and it is reference-data-economic as such data are only demanded where needed
in a rigorous, objective fashion whose accuracy only depends on the choice of the molecular-similarity measure. 
At first sight, this seems to come with the disadvantage that reference data must be provided on demand. However,
coupled cluster calculations can be easily launched for single-reference cases, and even multi-configurational
calculations can be launched in a fully automated fashion.\cite{Stein2016,Stein2016a,Stein2017,Stein2019}
To obtain accurate electronic structures and energies for molecules of increasing size, one may resort to embedding
approaches (see Refs. \citenum{Manby2012a,Muhlbach2018,Ye2020} and references therein) so that the quantum region which must be accurately described remains
to be of constant size.

It is important to appreciate that machine learning approaches can also take into account the residual error in the reference data
(see, e.g., Ref. \citenum{Proppe2019} for an application). Clearly, this only shifts the whole error-assessment problem to the model of higher accuracy:
it will require system-specific error estimation for the accurate reference, which will become increasingly
difficult for large molecules and therefore some sort of {\it error archigenesis} will be inevitable. 
As long as a rigorous error cannot be calculated
based on the structure of the underlying partial differential equations, highly accurate calculations on small
systems need to be carried out and assumptions on the residual error must be made. Then, such an axiomatic error may be
scaled up to the actual molecular size in an incremental scheme in which long-range correlations are sampled.

Here, it is implicitly assumed that the more reliable the error assessment is, the more demanding it will become in terms
of computational effort. It is for this reason that a hierarchical deployment of electronic structure models 
of different degree of reliability is a reasonable way to proceed in expansive computational campaigns
in order to accomplish reliable error estimation
in a systematic hierarchical manner.
At the level of error archigenesis, the sampling of long-range correlations is likely to be time consuming 
which will restrict the number of such calculations to be feasible in practice. Hence, a comparatively small number of
highly accurate reference results must be well chosen to connect to a vast number of results obtained with an
approximate model in order to assess its error for the individual molecules under consideration.

\begin{center}
{\bf Conclusion 2:}
{\it
 Bayesian uncertainty quantification is a way out of the error-assessment problem, 
but requires continuous benchmarking -- ideally with error assignment for the reference data themselves.
This benchmarking needs to be adjusted to the specific systems under study.
Reference data point calculations must be selected on the basis of confidence intervals taken, for instance,
from an underlying machine learning model.
Its accuracy will depend on the measure with which one strides across
the parameter space (e.g., the space of all molecular structures defined by nuclear scaffolds that parametrize a Born-Oppenheimer surface). 
}
\end{center}

\section{The Fate of Electronic Structure Models}

The historical development of electronic structure models has delivered many different classes of approximations with
different advantages and, hence, different areas of applications where they can blossom. Among these classes, variations
appear to be numerous and are typically characterized by capital-letter acronyms. Whereas only some of them are designed to
clearly replace older versions, most acronyms never die and are likely to survive in the future. The fact that very many
physical effects need to be approximated to eventually allow for a direct comparison with experiment (such as increasing molecular
sizes and complex environments (solvents, surfaces, enzymes, zeolites, metal organic frameworks, ...))
 together with the desire to propose new, potentially more successful methods has led
to an ever increasing number of acronyms. Already now, there are so many of them that it is difficult even for 
experts to keep track, and one wonders where this will lead to in a few decades because we may get lost
in a plethora of acronyms for variations of methods that may eventually choke scientific progress. 

By contrast, the discussion in the last section points to the possibility to stay with
a sufficiently reliable approximation of a generalist electronic structure model and to improve on it by selective and
directed refinement through system-specific benchmarking. Such a Bayesian approach supplements any approximate model with uncertainty
quantification for cases where no reference data is available and may even produce a system-focused variant of increased
accuracy. The system-specific uncertainty quantification alleviates the problem of 
questionable or unknown transferability of traditional benchmark sets and eventually allows us to make
better founded predictions on chemical processes.

This is clearly a compelling path forward, but it might render many models superfluous. For instance, if electronic energies
and molecular structures can be predicted with uncertainties assigned, only one rather than one hundred variations of some
type of approximation will be sufficient (such as one density functional at a given rung of Jacob's ladder
or even one for the whole ladder, for which a computationally efficient one can be chosen). 
This consequence is in accord with the conclusions drawn from the massive search for functionals by 
Mardirossian and Head-Gordon \cite{Mardirossian2014a,Mardirossian2017,Mardirossian2018} who showed that not a single 
unique functional would emerge, but very many functionals that differ only slightly in terms of accuracy measured at a large reference
data set. A slight change of this reference data set would then favor a different functional out of the pool of well-performing ones.

\begin{center}
{\bf Conclusion 3:}
{\it
One needs to know how accurate a computed result will be for a specific application and Bayesian
error estimation can be the key to provide this information, also weeding models of similar type and accuracy, hence
reducing the number of models that will be required.
}
\end{center}

\section{System-Focused First-Principles Models}

Consider the idea that benchmark sets will, in general, never be large enough for transferable parametrizations that are reliable
(even if a parametrized first-principles model features a high degree of flexibility).
Hence, reliable uncertainty quantification for a specific new molecular structure will uncover 
features that are not well represented in the already existing benchmark data
-- based on a suitable measure for molecular and electronic similarity that is key for the transferability assessment of already existing data.
Then, continuous benchmarking (i.e., benchmarking upon request) will be the only way out if results shall be supplemented by 
reliable Bayesian error
estimates. However, this comes with the benefit that one may re-adjust model parameters for a specific molecular
system to increase the accuracy of the approximate model in the local region of chemical space assigned by the similarity measure. 
Surely, one is advised to restrict such a reparametrization to values in the vicinity
of the literature values, provided that they define a suitable generalist model, in order to avoid a skew model that might not even
be transferable within a set of very similar molecular structures.

Hence, such a system-focused re-parametrization can increase the accuracy of an approximate electronic structure model 
in such a way that a subsequent $\Delta$-machine learning is likely to require less reference data because it may
be anticipated that it will be easier to learn differential rather than absolute effects, if the physical model
is suitable and flexible and if reference data are limited.

A system-focused model may be developed in the neighborhood of a well-established one.
Whether or not such a procedure is worthwhile will depend on the type of application. If only few calculations
are to be carried out, one is advised to choose a model that may be considered to have generally small errors (such as
a coupled cluster model). If, however, a massive computational campaign (as in reaction network explorations \cite{Sameera2016,Vazquez2018,Dewyer2018,Simm2019,Unsleber2020}
first-principles molecular dynamics studies,\cite{Marx2009} or high-throughput quantum chemical screening calculations \cite{Shoichet2004,Pyzer-Knapp2015}) is part
of solving a scientific problem, then a delicate balance of fast methods equipped with uncertainty quantification
through automated benchmarking upon request will be required. In order to limit the number of computationally expensive
calculations needed for the continuous production of reference data, one may exploit existing data of such a campaign to
refine the fast, but inaccurate model in a system-focused manner, as shown in Refs. \citenum{Brunken2020,Brunken2021}.  

If system-focused reparametrizations are carried out in the neighborhood of existing electronic structure models,
more accurate ab initio reference data can be used for this purpose so that the procedure does not introduce
additional empiricism, but may result in a potential loss of generality of the parent model.
However, this is made up for by the simultaneous assignment of uncertainty, for instance, through a machine learning model such as GPR or
a neural network representation that mediates between the fast-approximate results and the fewer accurate-but-expensive reference
points. Naturally, it will be beneficial to collect expensive reference data of known accuracy in globally
accessible data bases for permanent re-use.  

Requirements for a successful system-specific readjustment of parameters to decrease uncertainties are:
(1) reparametrization must be fast (i.e., no computational bottleneck must be created), 
(2) autonomy of the reparametrization must be established (i.e., no or hardly any human interference or input in order to be practical), and
as a consequence,
(3) full automation of all calculation steps must be guaranteed (especially for the
automated launch of reference calculations, which may require multi-configurational approaches).

\begin{center}
{\bf Conclusion 4:}
{\it
 Generalist electronic structure models cannot be accurate and fast at the same time (for large molecules or
molecular aggregates). A significant gain in computational
efficiency requires approximations that compromise transferability and accuracy. System-focused models  
can be made fast and accurate, but their transferability, i.e., their accuracy for related
structures, must be monitored
because they have not been tested on some arbitrary reference data. 
However, such reference data will be incomplete anyways if scrutinized with a suitable measure of molecular and electronic 
similarity.
Having system-focused models requires autonomy regarding parametrization and uncertainty quantification, and therefore, 
efficient automated procedures are needed in order to be practical.
}
\end{center}

\section{Outlook}

To know the systematic error -- within a well-defined confidence interval --
of a numerical result obtained from an approximate electronic structure model
is key for assessing the reliability and usefulness of a computational prediction in (bio)chemistry,
molecular physics, and materials science. 
It is risky, if not unacceptable, to rely on standard statistical measures such as the mean absolute deviation obtained for
some reference data whose relation to the case under study is not clear in a quantitative sense.
In this work, possible options to deal with this problem have been discussed. What has been elaborated here for
electronic-structure methods also extends to other branches of theoretical chemistry and to natural sciences 
in general (consider, for instance, vibrational structure problems with a Taylor series representation of
the electronic energy or strong-correlation methods in physics).

Rigorous Bayesian uncertainty quantification measures have been developed, but
having error estimation in place comes with an additional computational burden. Apart from the uncertainty quantification procedure
itself, a significant portion of the extra effort is due to the necessity of (continuous) benchmarking on demand. Still, this
should be embraced given that running reference calculations continuously is a way to produce system-specific error estimates in a well-defined objective manner
and a way to get rid of numerous variations of computational results for the same quantity
owing to the ever increasing number of electronic structure models.
It is critical for the reference data generated that also the uncertainty of the reference data is assessed and fed into the
overall error estimation process. Eventually, such a nesting procedure of methods ranging from fast but rather inaccurate
to accurate but computationally expensive approaches with decreasing uncertainty in the calculated result obtained for a specific
molecular system will require some sort of error archigenesis that estimates the error still immanent
in the most accurate approach available.

It is obvious that the accuracy of a calculated electronic structure result will not be sufficient for comparison with experiment. 
Additional effects need to be taken into account, most importantly those of molecular structure. In other words, 
nuclear motion and nuclear coordinates will affect the result that is eventually to be compared with an experimentally
measured quantity. Clearly, any additional computational modeling step will then also require rigorous uncertainty
quantification. As an example, we refer to the role of nuclear coordinates in spectroscopy that introduces uncertainties 
in the spectra discussed in Refs. \citenum{Oung2018,Bergmann2020}. A key challenge, that is likely to remain to a certain degree
at the discretion of the person conducting a computational study -- especially for large, nano-scale, composite atomistic systems --
is the choice of the atomistic configuration as a whole. If a selected atomistic structure does not represent the problem well,
no reliable uncertainty quantification will be able to detect this, and the calculated results, even when equipped with
rigorous error estimates, will not at all relate to the experimental situation (the assignment of protonation sites in a protein is only one
example that highlights the problem of choosing the proper molecular structure for a large system).

The ubiquitous availability of computing resources will make high-through\-put calculations more and more
routine. In such cases, very many calculations on rather similar molecular structures are to be performed. Examples
are again first-principles molecular dynamics simulations,\cite{Marx2009} high-throughput virtual property screenings,\cite{Shoichet2004,Pyzer-Knapp2015} and exploration of complex
reaction networks \cite{Sameera2016,Vazquez2018,Dewyer2018,Simm2019,Unsleber2020}. For these applications, it is clearly advantageous to employ
a fast approximate method (possibly refined in a system-focused manner) as often as possible and to restrict
resource-intensive benchmark calculations to a minimum. 

Having system-specific benchmark data available comes with the advantage that a system-focused model development
becomes easily possible through reparametrization of the fast model in the vicinity of the existing parameters of its
generalist version. This bears the option to improve on the accuracy of the fast model for further calculations 
and to save on additional expensive reference calculations. The connection between results obtained from
the many fast and few accurate calculations
can be made by data driven approaches. 
Obviously, to be feasible and routinely applicable, system-focused reparametrizations must be autonomous, which
requires fully automated procedures. 

Although computational resources are ubiquitous these days, the environmental footprint of quantum chemical calculations must be reduced. 
It is for this reason that also reference data supplemented with quality labels derived from uncertainty quantification 
should be stored in a globally accessible data base rather than being re-calculated locally.
Moreover, having nested computations of controlled accuracy covering a range from fast to expensive models
that is controlled by uncertainty quantification is a way to reduce the overall amount of atomistic simulations
required. Of course, this demands that uncertainty measures are of high reliability so that sufficiently accurate
predictions are obtained and clear-cut conclusions can be drawn. Such uncertainty-measure equipped results can also provide
a trustful basis for qualitative analysis, reasoning, and understanding in terms of chemical concepts.

\providecommand{\refin}[1]{\\ \textbf{Referenced in:} #1}

\end{document}